\documentclass[12pt,a4paper]{article}
\usepackage{graphicx}

\newcommand{\lesssim}{\,\raisebox{-.4ex}{%
\mbox{$\scriptstyle\stackrel{<}{%
\sim}$}}\,}

\begin{document}

\title{Assisted inflation in Friedmann-Robertson-Walker and Bianchi
spacetimes} 

\author{Juan M. Aguirregabiria, Alberto Chamorro, \\
{\it F\'{\i}sica Te\'orica, Universidad del Pa{\'\i}s Vasco,}\\ 
{\it Apdo.\ 644, 48080 Bilbao, Spain,}\\ 
Luis P. Chimento and Norberto A. Zuccal\'{a}\\ 
{\it Departamento de F\'{\i}sica, }\\ 
{\it Facultad de Ciencias Exactas y Naturales, }\\ 
{\it Universidad de Buenos Aires }\\ 
{\it Ciudad Universitaria, Pabell\'{o}n I, }\\ 
{\it 1428 Buenos Aires, Argentina.}}
\maketitle

\begin{abstract}
We use exact general solutions for the spatially flat FRW and the
anisotropic Bianchi~I cosmologies to show that generically uncoupled
scalar fields cooperate to make inflation more probable, while the
presence of several interacting fields hinders the occurrence of the
phenomenon, in accordance with previous results based on particular
power-law solutions. Similar conclusions are reached in the case of
Bianchi~VI$_0$ spacetimes, for power-law solutions which are proved to be
attractors. \end{abstract}

\vspace*{1cm} \noindent PACS 04.20.Jb



\newpage


\section{Introduction}

\label{sec:intro}

In many inflationary models the effective potential energy density of a
scalar field is responsible for an epoch of accelerated inflationary
expansion~\cite{inflation}. Very often one assumes that inflation is
driven by a scalar field of the Liouville form, i.e., a exponential
potential, because this kind of potential arises in various
higher-dimensional supergravity~\cite{salam} and
superstring~\cite{fradkin} models~\cite{olive,taha,easther,Wands}.

Although there are many scalar fields in superstring theories, in the
past it was often assumed that typically only one scalar field was
responsible for the inflation, while those having higher exponents were
quickly redshifted away. However, it has been found~\cite{lms} that the
so-called {\em assisted inflation} may occur when several scalar fields
are present, even if each individual field is too steep to drive the
inflation, provided that the fields are uncoupled and interact only
through the geometry. On the other hand, if the fields interact directly
with each other, the opposite effect may happen and the presence of
cross couplings beteween fields may hinder
inflation~\cite{ko,hawk,copeland}.

Assisted inflation has been mainly studied in power-law solutions
($a\propto t^p$) for the spatially flat Friedmann-Robertson-Walker (FRW)
cosmology, which can be shown~\cite{lms} to be the late-time attractor
for the evolution of this kind of model. (Recently, Green and
Lidsey~\cite{green} discussed in the context of assisted inflation the
late-time evolution in a general geometry.)

The purpose of this work is to extend previous studies on multi-scalar
field cosmologies in two directions: firstly, we will use general
solutions (instead of the special power-law ones) to analyze if the
presence of several fields generically helps or impedes inflation, and
secondly, going beyond the aforementioned FRW cosmology, we will
consider the anisotropic inhomogeneous generalization given by Bianchi
type I models. Thus in section~ \ref{sec:intfrw} we deal with $n$
interacting scalar fields in a FRW spacetime and use the general
solution to show that the larger is the number of interacting scalar
fields the less likely is inflation. Section~\ref {sec:intspace} makes
plausible for more general scalar field potentials the results obtained
in section~\ref{sec:intfrw} for exponential potentials. We do not know
the solution for $n$ non-interacting scalar fields in a FRW cosmology in
the general case, but we use the discussion of the late-time attractor
of section~\ref{sec:attractor} to restrict the analysis of uncoupled
fields in section~\ref{sec:nonintfrw} to the particular case in which
all fields are assumed to be equal, for which the general solution can
be found. We show that non-interacting fields generically cooperate to
assist inflation. The density fluctuations corresponding to the last
case are discussed in section~\ref{sec:density}. General solutions of
anisotropic Bianchi~I cosmologies with interacting and uncoupled fields
are used in the first part of section~\ref{sec:intbianchi} and
subsection~\ref{sec:nonintbianchi}, respectively, to check that also in
those cases interacting fields make inflation more difficult while
uncoupled fields assit it. The stability of power-law solutions is
dicussed in~\ref{sec:stabilityI}. Finally in section~ \ref{sec:bianchi6}
we turn to power-law solutions of the Bianchi~VI$_0$ model, reaching
again the same conclusions.


\section{The $n$-scalar field problem in a flat FRW spacetime}

\label{sec:FRW}

In the following we will consider two kinds of problems in flat FRW
spacetimes in which there are $n$ homogeneous scalar fields driven by
exponential potentials. First of all, we will assume that the scalar
fields are interacting through a product of exponential potentials. Then
we will consider the case in which the scalar fields are uncoupled 
because
the potential is a sum of potentials involving a single field.


\subsection{The interacting $n$-scalar field problem in flat FRW
spacetime}

\label{sec:intfrw}

The problem of $n$ interacting homogeneous scalar fields, $\phi_i$, 
driven by a product of $n$ exponential potentials $V_i=V_{0i}{\rm
e}^{-k_i\phi_i}$ minimally coupled to gravity in a flat Robertson-Walker
spacetime, with metric 
\begin{equation}  \label{rw}
ds^{2}=-dt^{2}+a^{2}(t)\left[dx^{2}+dy^{2}+dz^{2}\right], 
\end{equation}
is formulated by the system of equations 
\begin{equation}  \label{00}
3H^{2}=\frac{1}{2}\dot\phi^{2}+V, 
\end{equation} 
\begin{equation} 
\label{kg} 
\ddot{\vec\phi}+3H\dot{\vec\phi}-V\vec k=0, 
\end{equation}
where $H=\dot a/a$ and the potential 
\begin{equation}  \label{po}
V(\vec\phi)=V_{0}{\rm e}^{-\vec k\cdot\vec\phi}, 
\end{equation}
allows
for interactions between the fields. $V_0$ is the constant $
V_{01}V_{02}\cdots V_{0n}$, and $\vec k=(k_1,k_2,...,k_n)$ is an $n$
-component constant vector with respect to an orthonormal basis in the
$n$ -dimensional Euclidean internal space to which the vector $\vec\phi
=(\phi_1,\phi_2,...,\phi_n)$, built with the $n$ minimally coupled
scalar fields, also belongs. In the remaining of this section $k$, $k^2$
and $ \dot\phi^2$ stand for $\left|\vec k\right|$, $\vec k\cdot\vec k$,
and $\dot{ \vec\phi}\cdot\dot{\vec\phi}$ respectively. Potentials of
this type are of interest because they may be considered just as an
approximation to a more complex potential. In fact, in
higher-dimensional superstring theories, the scalar field is like one of
the matter fields that contribute to the action and effective potential
of the theory. Loop expansion~\cite{fradkin}, or expansion in the number
of interacting particles~\cite{gross}, of the action leads to a
perturbative expression of the potential which is a summation of
exponential terms~\cite{{taha},{easther}}. From (\ref{00})--(\ref{kg}),
and discarding the trivial static metric solution $H=0$,
$\frac12\dot\phi^2+V=0$ , we get 
\begin{equation}  \label{hp}
\dot{H}=-\frac{1}{2}\dot{\phi}^{2}. 
\end{equation}
Using this equation
and the system (\ref{00})--(\ref{kg}) one finds that the Klein-Gordon
equations (\ref{kg}) have the first integrals 
\begin{equation}
\label{pi} \dot{\vec\phi} =H\vec k+\frac{\vec c}{a^{3}}, 
\end{equation}
where $\vec c=(c_1,c_2,...,c_n)$ is an arbitrary vector integration
constant. As (\ref{pi}) involves only geometrical quantities the
Einstein-Klein-Gordon equations uncouple and the general parametric
solution can be obtained~\cite{jm},~\cite{nor},~\cite{2c}.

One can easily verify that in this kind of models the
Einstein-Klein-Gordon equations have power-law solutions 
\begin{equation}
a\propto t^{\frac{2}{k^2}},  \label{eq:powerlaw} 
\end{equation} 
so that they inflate at all times when $k^2<2$. We will show below that
the general solution has the same kind of behavior when the scale factor
$a$ is large enough.

Inserting (\ref{pi}) in (\ref{hp}) we obtain the following second-order
equation for the scale factor $a(t)$ 
\begin{equation}  \label{eq:sequ}
\ddot{s}+s^{m}\dot{s}+\frac{1}{4\cos^2\sigma}\,\,s^{2m+1}=0, \qquad
(\sigma\ne \pi/2), 
\end{equation}
where 
\begin{equation}
\cos\sigma=\frac{\vec c\cdot\vec k}{c k},  \label{si} 
\end{equation}
$m=-6/k^2<0$, $c=\left|\vec c\right|$, $a$ and $t$ have been replaced by
the new variables $s$ and $\tau$ defined by 
\begin{equation}  \label{a}
a=s^{-\frac{m}{3}}, \qquad \tau=ckt\cos\sigma, 
\end{equation}
and the dot
stands for derivation with respect to $\tau$. Once $s(\tau)$ is known one
can compute, in principle, the scale factor $a(\tau)$ from (\ref{a} ),
and the fields ${\vec \phi(t)}$ from equations (\ref{pi}).

Equation (\ref{eq:sequ}) is a particular case of the second order
nonlinear ordinary differential equation 
\begin{equation}  \label{f}
\ddot{s}+\alpha f(s)\dot{s}+\beta f(s)\int{f(s)\,ds}+ \gamma f(s)=0,
\end{equation}
where $f(s)$ is some real function and $\alpha$, $\beta$
and $\gamma$ are constant parameters. Depending on the values of $k^2$ our
equation~(\ref {eq:sequ}) corresponds to the next two cases:
\begin{enumerate}
\item  $k^2\ne 6$ and 
\begin{equation}
f(s)=s^m,\qquad \alpha =1,\qquad \beta =\frac{m+1}{4\cos ^2\sigma },\qquad
\gamma =0.  \label{151} 
\end{equation}
\item  $k^2=6$ and 
\begin{equation}
f(s)=\frac 1s,\qquad \alpha =1,\qquad \beta =0,\qquad \gamma =\frac
1{4\cos ^2\sigma }.  \label{17} 
\end{equation}
\end{enumerate}

Inserting the first integrals (\ref{pi}) in the Einstein equation
(\ref{00} ), we get a quadratic equation in the expansion rate $H$ which
has real solutions only when its discriminant is non-negative. This
condition leads to 
\begin{equation}  \label{3.14}
\beta\le\frac{1}{4\left[1+\frac{2a^6V}{c^2}\right]}<\frac{1}{4},
\end{equation}
discarding the oscillatory solutions of equation~(\ref{f}),
which would be obtained for a negative potential or $\beta> 1/4$. It
follows that, if $ k^2\ne6$, real solutions exist always for
\begin{equation}  \label{3.15} k^2<k^2_0=\frac{6}{\sin^2\sigma}.
\end{equation}
This sets a restriction on the integration constants and
the exponent of the exponential-potential.

The general solution of (\ref{f}) can be obtained making the nonlocal
transformation of variables~\cite{lmath} 
\begin{equation}  \label{3.16}
z=\int{f(s)\,ds}, \qquad \eta=\int{f(s)\,d\tau}. 
\end{equation}
Under this
transformation (\ref{eq:sequ}) becomes a linear inhomogeneous ordinary
differential equation with constant coefficients 
\begin{equation} 
\label{3.17} z^{\prime\prime}+\alpha z^{\prime}+\beta z +\gamma=0,
\end{equation}
which for $\alpha>0$ and $\beta>0$ is the equation of a
damped harmonic oscillator in a constant external field. Here the
$^{\prime}$ indicates differentiation with respect to $\eta$. In our case
the nonlocal transformation of variables (\ref{3.16}) is
\begin{equation}  \label{5.1}
z=\frac{s^{m+1}}{m+1}, \qquad \eta=\int{s^m\,d\tau},
\end{equation}
for $k^2\ne6$, and 
\begin{equation}  \label{5.11}
z=\ln s, \qquad \eta=\int{\frac{\,d\tau}{s}},
\end{equation}
for $k^2=6$.

Taking $\alpha=1$ in (\ref{3.17}) we get 
\begin{equation}
z=\left\{ 
\begin{array}{ll}
b_1{\rm e}^{\lambda_+\eta}+b_2{\rm e}^{\lambda_-\eta},\qquad & \mbox{for
} k^2\ne 6; \\ b_1+b_2{\rm e}^{-\eta}-\gamma\eta,\qquad & \mbox{for
}k^2=6; \end{array} \right.  \label{5.3} 
\end{equation}
where
\begin{equation}  \label{5.4} \lambda_\pm=\frac{-1\pm\sqrt{1-4\beta}}{2},
\end{equation}
and $b_1$, $b_2$ are real integration constants.

We see from~(\ref{a}) and~(\ref{5.1}) that 
\begin{equation}
a\propto z^{-\frac{m}{3(m+1)}},  \label{eqaprop}
\end{equation}
so that, while for $k^2>6$ the scale factor goes to $0$ as $z\to0$, one
has that for $k^2<6$ the scale factor goes to infinity when $z\to0$. On
account of equations~(\ref{5.3}) one may assume that the latter occurs as
$\eta\to\eta_0$, for an appropriate value of $\eta_0$. Let us now write
$\eta=\eta_0+\delta\eta$ and expand (\ref{5.3}) and~(\ref{eqaprop}) about
$z=0$ and $\eta=\eta_0$, keeping only terms linear in $\delta\eta$, so
that $\delta z =\delta \eta$ and 
\begin{equation} a=\delta
z^{-\frac{m}{3(m+1)}}= \delta\eta^{-\frac{m}{3(m+1)}}. \label{eq:adelta}
\end{equation}
Now, we see from (\ref{a}) and~(\ref{5.1}) that
\begin{equation}  \label{de} \delta\eta=a^{-3}\delta\tau\propto
a^{-3}\delta t. 
\end{equation}
If we use this in (\ref{eq:adelta}), we get
successively 
\begin{equation} a\propto a^{\frac{m}{m+1}}\,{\delta
t}^{-\frac{m}{3(m+1)}}, 
\end{equation}
and 
\begin{equation} a\propto
{\delta t}^{-\frac{m}{3}}={\delta t}^{\frac{2}{k^2}}. 
\end{equation}
We
see then that if $k^2<2$ the solution does inflate at least along some
time interval. Now, since $k^2=k_1^2+k_2^2+\cdots+k_n^2$, one concludes
from our analysis of the general solution that the larger is the number of
the interacting scalar fields the less likely will be $k^2<2$ as well that
inflation take place. On the other hand, for $6<k^2<k_0^2$, one has $
0<\beta<1/4$ that implies that both $\lambda_{\pm}<0$; thus (\ref{5.1})
and~( \ref{5.3}) tell us that the scale factor goes to infinity when
$z\to\infty$ and $\eta\to -\infty$ (with no loss of generality we are
taking both constants $b_1, b_2 > 0$). By keeping only the dominant terms
we have for very large negative $\eta$ that 
\begin{equation} a\propto {\rm
e}^{-\frac{m\lambda_-\eta}{3(m+1)}}. 
\end{equation}
This relation and
(\ref{de}), that is generally valid, yield 
\begin{equation} 
\label{eq:deltaaa} \frac{\delta a}{a}\propto a^{-3}\delta t 
\end{equation}
whose general solution is $a\propto t^{1/3}$. In this case the solution
does not inflate and it approaches to the free scalar field solution.
These results are in accordance with those obtained with particular
solutions by other authors~\cite{ko,hawk,copeland}.

It would be surprising if for $k^2=6$ the scale factor had a behavior
drastically departing from that suggested by the previous analysis when $
k^2\in\left(k_0^2,6\right)\bigcup(6,2)$. In fact, we see from (\ref{a})
and (\ref{5.11}) that $z=3\log a$ so that, as a consequence of
(\ref{5.3}), $a$ goes to infinity as $\eta\to-\infty$. From (\ref{5.3}) we
see that $\log z \sim -\eta$, when $\eta\to-\infty$
and, by using (\ref{de}),
 \begin{equation}
\frac{\delta a}{a\log a}\propto\frac{\delta
z}{z}\sim-\delta\eta\propto\frac{\delta t}{a^3},
 \end{equation} 
which can be written as
 \begin{equation}\label{eq:log}
\frac{\delta a}{a}\sim a^{-3}\delta t\,\log a.
 \end{equation} 
In the asymptotic regime in which $a\to\infty$ we have $a\,\delta a<
a^2\,\delta a/\log a < a^2\, \delta a$, so that from (\ref{eq:log}) we get
$a^2\lesssim t \lesssim a^3$ and, finally,
  \begin{equation}
  t^{1/3}\lesssim a \lesssim t^{1/2},\qquad\mbox{when }a\to\infty.
 \end{equation}
We see that the solution does not inflate.


\subsection{More general potentials}

\label{sec:intspace}

The results of the above section show that we can introduce an $n$
-dimensional Euclidean internal vector space containing the $n$-component
vector $\vec\phi =(\phi_1,\phi_2,...,\phi_n)$ built with $n$ minimally
coupled scalar fields. Let us assume now that the potential has the
general form 
\begin{equation}  \label{vin} V=V(\Phi), \qquad \Phi=\vec
k\cdot\vec\phi. 
\end{equation}
In this case the Einstein equation
(\ref{00}) remains unchanged, however the Klein-Gordon equation (\ref{kg})
becomes 
\begin{equation}  \label{kgin}
\ddot{\vec\phi}+3H\dot{\vec\phi}+V^{\prime}\vec k=0. 
\end{equation}
where
the $^{\prime}$ indicates derivatives with respect to the variable $
\Phi$. Since (\ref{00}), (\ref{vin}) and (\ref{kgin}) are invariant under
rotations of the orthogonal axes in the Euclidean space above mentioned,
we may choose the first axis of this internal space along the vector $\vec
k$. Then the Klein-Gordon equation splits into one equation for
$\phi_1=\phi$ 
\begin{equation}  \label{kgp}
\ddot\phi+3H\dot\phi+kV^{\prime}=0, 
\end{equation}
and $n-1$ free field
Klein-Gordon equations for $\phi_2=\cdots=\phi_n=\psi$ 
\begin{equation} 
\label{kgs} \ddot\psi+3H\dot\psi=0. 
\end{equation}
From (\ref{kgs}) we
obtain the first integral $\dot\psi=c_0/a^3$, where $ c_0 $ is an
arbitrary integration constant. Hence the original $n$-scalar field
problem is equivalent to considering a self-interacting scalar field with
stiff matter. In fact, the Einstein equation (\ref{00}) now reads
\begin{equation}  \label{00in}
3H^{2}=\frac{1}{2}\dot\phi^{2}+V+\frac{c_\psi^2}{2a^6}, 
\end{equation}
where $c_\psi^2$ is the sum of $n-1$ positive defined integration
constants.

Taking into account that the scalar field $\phi$ depends only on $t$, its
energy-momentum tensor may be written in the perfect fluid form
\begin{equation}  \label{Tfluid} T_{ik}= (p_{\phi}+\rho_{\phi})
u_{i}u_{k}+p_{\phi}g_{ik}\,, 
\end{equation}
where \begin{eqnarray} 
\label{rhophi} \rho_{\phi} & = & \frac{1}{2} \dot\phi^2 + V(\phi)\, , 
\nonumber \\ p_{\phi} & = & \frac{1}{2} \dot\phi^2 - V(\phi).
\end{eqnarray} The fluid interpretation of the scalar field has proven
very useful in the study of the inflationary and Q-matter scenarios
\cite{cjp}. In particular it leads to consider its equation of state
$p_{\phi}=\left(\gamma_{\phi}-1 \right)\rho_{\phi}$. On the other hand,
the state equation for fluid representing stiff matter is
$p_f=\left(\gamma_f-1\right)\rho_f$ with $\gamma_{f}=2$. Because of the
additivity of the stress-energy tensor it makes sense to consider an
effective perfect fluid description with equation of state
$p=\left(\gamma-1\right)\rho$ where $p=p_f+p_{\phi}$, $
\rho=\rho_f+\rho_{\phi}$ and 
\begin{equation}  \label{ga}
\gamma=\frac{\gamma_f\rho_f+\gamma_{\phi}\rho_{\phi}}{\rho_f+\rho_{\phi}}
\end{equation}
is the overall (i.e. effective) adiabatic index. For this
effective perfect fluid the dynamical equations are 
\begin{equation}
3\frac{\dot a^2}{a^2}=\rho 
\end{equation}
and 
\begin{equation} 
\label{a11} \frac{\ddot a}{a}=-\frac{1}6\left[\rho+3p\right],
\end{equation}
where $p$ and $\rho $ are the density and pressure of the
effective perfect fluid. They involve the self-interacting scalar field
and the $n$-1 free scalar fields. Inflationary solutions occur when $\ddot
a>0$; this means that the expansion is dominated by a gravitationally
repulsive stress that violates the strong energy condition $\rho+3p<0$ or
equivalently $\gamma<2/3$ . When we impose this condition on (\ref{ga}),
we obtain that 
\begin{equation}  \label{vin>}
V(\Phi)>\dot\phi^2+\frac{c_\psi^2}{a^6}. 
\end{equation}
Hence, when there
is only one scalar field the solution inflates for $ V(\phi)>\dot\phi^2$.
However, the $n$ interacting scalar fields produce a desassisted inflation
because the solutions inflate only if (\ref{vin>}) holds. Using the
effective perfect fluid description we can generalize this analysis to an
arbitrary potential $V(\phi_1,\phi_2,...,\phi_n)$. In this case we obtain
that $V(\phi_1,\phi_2,...,\phi_n)>\dot\phi_1^2+\dot
\phi_2^2+.....+\dot\phi_n^2$, so that the interacting $n$ scalar fields
might make inflation more unlikely in a FRW spacetime.


\subsection{The $n$-scalar field attractor problem in the FRW spacetime}

\label{sec:attractor}

Let us now assume that the $n$ homogeneous scalar fields, $\phi_i$, in the
FRW spacetime are driven by a general potential $V=V(\phi_i)$. In that
case the Einstein-Klein-Gordon equations are  
\begin{equation}
\label{b001} 3H^{2}= \frac{1}{2}\sum_{i=1}^n \dot\phi_i^{2}+V,
\end{equation}  
\begin{equation}  \label{bkg1}
\ddot{\phi_i}+3H\dot{\phi_i}+V_{,\phi_i}=0. 
\end{equation}
 where
$V_{,\phi_i}$ stand for $\partial V/\partial {\phi_i}$. From these
equations we get  
\begin{equation}  \label{bhp}
\dot{H}=-\frac{1}{2}\sum_{i=1}^n \dot{\phi_i}^{2}. 
\end{equation}

In order to investigate the stable scalar field configurations, we
introduce the quantity  
\begin{equation} \label{om}
\omega=\frac{\sum_{i=1}^n \dot{\phi_i}^{2}}{n\dot\phi_{\alpha}^2},
\end{equation}
 which reduces to $\omega=1$ for the configuration
$\phi_1=\phi_2=\cdots=\phi_n$.  Using (\ref{b001})--(\ref{om}) we find the
differential equation for $\omega$:  
\begin{equation} \label{dom}
\dot\omega=2\frac{nV_{,\phi_\alpha}\dot\phi_{\alpha}\omega-\dot
V}{n\dot\phi^2_{\alpha}}. 
\end{equation}
 (In this section no summation
convention applies to repeated Greek indexes.) If we further assume that
the potential satisfies the condition  
\begin{equation} \label{poa} \dot
V=nV_{,\phi_\alpha}\dot\phi_{\alpha}, 
\end{equation}
 equation
(\ref{dom}) becomes  
\begin{equation} \label{doma}
\dot\omega=2\frac{V_{,\phi_\alpha}}{\dot\phi_{\alpha}}(\omega-1)
\end{equation}
 which has a fixed point solution: $\omega=1$.
Furthermore, the general solution of (\ref{doma}) can be found using
(\ref{bkg1}):  
\begin{equation} \label{oma}
\omega=1+\frac{c}{a^6\dot\phi^2_{\alpha}} 
\end{equation}
 where $c$ is an
arbitrary integration constant. However, it is useful to express the last
solution in terms of geometrical quantities with the aid of (\ref{bhp})
and (\ref{om}). The final result is  
\begin{equation} \label{omg}
\omega=\left(1+\frac{nc}{2a^6\dot H}\right)^{-1}. 
\end{equation}
Evaluating (\ref{omg}) in the asymptotic regime, it can be easily shown
that the particular solution $\omega=1$ is an attractor for evolutions
that behave asymptotically as $a\propto t^\nu$ with $\nu>1/3$.

This result strongly suggests that the special case in which all scalar
fields are equal may be the late-time attractor of more general problems.
In fact, this has been proved in \cite{ko} for  potential~(\ref{po}) with
$k_1=\cdots=k_n$ and $V_{01}=\cdots=V_{0n}$, which guarantee
assumption~(\ref{poa}). We did not use this fact in the previous section
because we were able to use the general solution with no additional
assumption on $k_i$, $V_{0i}$ and $\phi_i$, but it will be useful to
simplify somewhat the problem analyzed in the following section.


\subsection{The non-interacting $n$-scalar field problem in the FRW
spacetime}

\label{sec:nonintfrw}

Let us now assume that the $n$ homogeneous scalar fields, $\phi_i$, in the
spacetime given by~(\ref{rw}) do not interact directly, but are driven by
a sum of $n$ exponential potentials $V_i=V_{0i}{\rm e}^{-k_i\phi_i}$. In
that case the Einstein-Klein-Gordon equations are 
\begin{equation} 
\label{001}
3H^{2}=\sum_{i=1}^n{\left[\frac{1}{2}\dot\phi_i^{2}+V_i\right]},
\end{equation} 
\begin{equation}  \label{kg1}
\ddot{\phi_i}+3H\dot{\phi_i}-k_iV_i=0. 
\end{equation}

From now on we will consider a simplified problem in which 
$k_1=k_2=\cdots=k_n\equiv k$ and $V_{01}=V_{02}=\cdots=V_{0n}\equiv V_0$.
As discussed in the previous section, we can expect in this particular
case that in the asymptotic evolution all scalar fields tend to a common
limit. That this is actually the case has been proved in \cite{lms}. In
consequence, in the remaining of this section we will take
 $\phi_1=\phi_2=\cdots=\phi_n\equiv \phi$, so that
 $V_1=V_2=\cdots=V_n\equiv V=V_0{\rm e}^{-k\phi}$
and
equations~(\ref{001})--(\ref{kg1}) reduce to 
\begin{equation}  \label{002}
3H^{2}=\frac{n}{2}\dot\phi^{2}+nV,
\end{equation}
\begin{equation}  \label{kg2}
\ddot{\phi}+3H\dot{\phi}-kV=0.
\end{equation}
One can easily get from (\ref{002})--(\ref{kg2}) 
\begin{equation}  \label{hp1}
\dot{H}=-\frac{n}{2}\dot{\phi}^{2}
\end{equation}
and the first integral of the Klein-Gordon equation (\ref{kg2}) 
\begin{equation}  \label{pi1}
\dot{\phi} =\frac{k}{n}H+\frac{c}{a^{3}},
\end{equation}
where $c$ is an arbitrary integration constant.

As in the previous section, the Einstein-Klein-Gordon equations have
power-law solutions~\cite{lms} 
\begin{equation}
a\propto t^{\frac{2n}{k^2}},  \label{eq:powerlaw1}
\end{equation}
so that this type of solutions inflates at all times when $k^2<2n$. We
will now show that the general solution of the Einstein-Klein-Gordon
equations ( \ref{002}) and~(\ref{kg2}) has the same kind of behavior when
the scale factor $a$ is large enough.

Inserting (\ref{pi1}) in (\ref{hp1}) we obtain the following second-order
equation for the scale factor $a(t)$ 
\begin{equation}  \label{eq:sequ1}
\ddot{s}+s^{m}\dot{s}+\frac{1}{4}\,\,s^{2m+1}=0, 
\end{equation}
where
$m=-6n/k^2<0$, the dot means derivative with respect to $\tau$ and we have
used, instead of $a$ and $t$, the new variables $s$ and $\tau$ defined by
\begin{equation}  \label{a1} a=s^{-\frac{m}{3}}, \qquad \tau=c kt.
\end{equation}

We have again a particular case of equation~(\ref{f}) with $\alpha=1$ and
to find the general exact solution of (\ref{eq:sequ1}) one has to consider
the following two possibilities:

\begin{enumerate}
\item  $k^2\ne 6n$ and 
\begin{equation}
f(s)=s^m,\qquad \beta =\frac{m+1}4,\qquad \gamma =0.  \label{1512}
\end{equation}

\item  $k^2=6n$ and 
\begin{equation}
f(s)=\frac 1s,\qquad \beta =0,\qquad \gamma =\frac 14.  \label{172}
\end{equation}
\end{enumerate}

The general solution of (\ref{eq:sequ1}) can be obtained by performing the
nonlocal transformation of variables given in~(\ref{5.1})--(\ref{5.11})
with the new value of $m$, which reduces (\ref{eq:sequ1}) to the linear
inhomogeneous ordinary differential equation with constant coefficients
(\ref {3.17}).

If one now repeats the analysis of the previous section by systematically
taking $\sigma=0$ and using the new $m$, the same final results for $z$
and $a$ are obtained: just replace $k^2=k_1^2+\cdots k_n^2$ by $k^2/n$ in
the quantities involved in (\ref{5.3}) and (\ref{eqaprop}). One readily
concludes that, when $a\to\infty$, these models inflate if $k^2<2n$, so
that the fields cooperate to make inflation more likely in the so-called
``assisted inflation'', which was first discussed ---but only for
power-law solutions--- in~\cite{lms}.

This result could have been anticipated since the present case is included
in the mathematical problem set in section~\ref{sec:intfrw} by
equations~(\ref{eq:sequ}) and~(\ref{si}) by just taking $\cos \sigma
=\pm 1$ (that corresponds to the simplified problem under consideration)
and $m=-6n/k^2$.


\subsection{Density fluctuations}

\label{sec:density}

The fact that the contributions of the density fluctuations differ
significantly in different inflationary universe models, has motivated a
detailed study of all the alternatives. In this context, it is interesting
to derive the spectral indices for the perturbations that would be created
during the periods of inflation described by the solutions given in the
last section. It is well known that, for multi-scalar field models, the
spectrum of the curvature perturbation reads~\cite{Sasaki},
\begin{equation}
P_S=\left( \frac H{2\pi }\right) ^2\,\frac{\partial N}{\partial \phi
_i}\, \frac{\partial N}{\partial \phi _j}\,\delta _{ij},  \label{PR}
\end{equation}
\noindent where $N$ is the number of $e$-foldings of
inflationary expansion remaining, and there is a summation over $i$ and
$j.$ In the case considered in the previous section,where all the scalar
fields are equal, (\ref{PR}) yields \cite{lms} 
\begin{equation}
P_S({\tilde{k}})=\left. \left( \frac H{2\pi }\right) ^2\,\frac
1n\,\frac{H^2 }{\dot{\phi}^2}\right\rfloor _{aH={\tilde{k}}}.
\label{spectrum} 
\end{equation}
\noindent where $H$ and $\dot{\phi}$ have
to be evaluated at the time when the wave number of interest ${\tilde{k}}$
leaves the horizon during inflation. Also in this case, the spectral index
$n_S({\tilde{k}})$ defined as \[ n_S({\tilde{k}})=1+\frac{{\rm d}\,\ln
P_S}{{\rm d}\,\ln {\tilde{k}}} \] \noindent is given by \cite{Sasaki,lms}
\begin{equation} 1-n_S=-2\,\frac{\dot{H}}{H^2}.  \label{ns} 
\end{equation}
\noindent The availability of exact solutions allows us to express the
relevant quantities as functions of the variable $\eta $ introduced in
equation~(\ref{3.16}). The scale factor reads (see (\ref{eqaprop}), which
is  satisfied for uncoupled scalar fields with $m=-6n/k^2$, as one can
easily see) 
\begin{equation} a(\eta )=\left[ \left( m+1\right) \ z(\eta
)\right] ^{-\frac m{3\ \left( m+1\right) }}  \label{a(eta)} 
\end{equation}
where $z(\eta )$ is given by (\ref{5.3})--(\ref{5.4}). The spectrum
$P_S({ \tilde{k}})$ obtained from (\ref{spectrum}) is shown in Figure~1,
for $k=3$, $c=-1$, $b_1=-1$, $b_2=0.001$ and different values of $n$.
Here, inflation is possible for $n\geq 5$, and one can see that the peak
of the spectral distribution moves towards the high frequency region as
$n$ increases. The corresponding spectral index $n_S$ is shown in Figure
2. From (\ref{ns}) and the general solutions (\ref{5.3}), (\ref{5.4}) and
(\ref{a(eta)}), it can be shown that the value of $n_S\to1$ in the
asymptotic region $a\gg 1$ as the number $n$ of present fields increases.
This feature is exemplified in Figure~2, where we can see how the larger
is the value of $n$, the closer is the spectrum to the scale invariance
\cite{lms}.


\section{The $n$-scalar field problem in Bianchi
type I models}

\label{sec:Bianchi}

Now we turn to the general Bianchi type I model with
$n$ homogeneous scalar fields driven by
exponential potentials. As we proceeded in the case of FRW spacetimes, we
will first assume that the scalar fields are interacting through a product
of exponential potentials, an then we will consider the case in which the
scalar fields are uncoupled because the potential is a sum of potentials
involving a single field.


\subsection{The interacting $n$-scalar field problem in the aniso\-tropic
Bianchi type I model}

\label{sec:intbianchi}

The general Bianchi type I model is the anisotropic generalization of the
spatially flat FRW universe expanding differently in the $x$, $y$, and $z$
directions. In the usual synchronous form its line element is given by
\begin{equation} ds^2=-dT^2+a_1^2(T)\,dx^2+a_2^2(T)\,dy^2+a_3^2(T)\,dz^2. 
\label{metricz} 
\end{equation}
For convenience we use the semiconformal
coordinates 
\begin{equation} dt\equiv\frac{dT}{a_3},\quad {\rm e}^f\equiv
a_3^2,\quad G\equiv a_1a_2, \qquad {\rm e}^p\equiv\frac{a_1}{a_2},
\end{equation}
to cast the metric~(\ref{metricz}) into the form
\begin{equation} ds^2 = {\rm e}^{f(t)} \left(-dt^2+dz^2\right) + G(t)
\left({\rm e} ^{p(t)}\,dx^2 +{\rm e}^{-p(t)}\,dy^2\right).
\label{metric} 
\end{equation}

We first consider, as in section~\ref{sec:intfrw}, $n$ scalar fields
$\phi_i$ interacting directly through the exponential potential
(\ref{po}). The problem of $n$ interacting homogeneous scalar fields,
$\phi_i$, driven by a product of $n$ exponential potentials
$V_i=V_{0i}e^{-k_{i}\phi_{i}}$, minimally coupled to gravity in the
Bianchi~I spacetime (\ref{metric}), is formulated by the following system
of Einstein-Klein-Gordon equations 
\begin{equation} \dot p=\frac{a}{G}, 
\label{pression} 
\end{equation} 
\begin{equation} {\rm e}^f=\frac{\ddot
G}{2VG},  \label{ef} 
\end{equation} 
\begin{equation} \frac{\ddot
G}{G}-\frac{1}{2}\left(\frac{\dot G}{G}\right)^2 - \frac{\dot G}{ G}
\dot{f} + \frac{1}{2}\dot p^2 = - \dot\phi^2,  \label{G2} 
\end{equation}
\begin{equation}  \label{klg} \ddot{\vec\phi}+\frac{\dot
G}{G}\dot{\vec\phi}-{\rm e}^f\,V\vec k=0, 
\end{equation}
where $a$ is an
arbitrary integration constant. It can be easily seen that the vector
\begin{equation}  \label{2} \dot{\vec\phi} =\frac{\dot G}{G}\frac{\vec
k}{2}+\frac{\vec m}{G}, 
\end{equation}
(where $\vec m$ is a
$n$-dimensional vector whose components are integration constants) is a
first integral of the Klein-Gordon equation set (\ref{klg}). Inserting
(\ref{pression}) in equation~(\ref{2}) the general solution of the
Klein-Gordon equations is found: 
\begin{equation}  \label{field}
{\vec\phi}={\vec\phi_0}+p\frac{\vec m}{a}+\frac{\vec k}{2}\ln G,
\end{equation}
where $\vec\phi_0$ is an arbitrary constant vector.
Equations (\ref{pression} )--(\ref{G2}) along with (\ref{field}) uncouple
and their solutions can be obtained if one is able to solve the following
third-order equation for $G$ 
\begin{equation} G\ddot
G^2-{\buildrel\ldots\over G}\dot G G +\left(\frac{1}{2}- \frac{k^2
}{4}\right) \ddot G\dot G^2 + \left(m^2 +\frac{a^2}{2}\right)\ddot G = 0.
\label{G3} 
\end{equation}
Once $G(t)$ is known, in principle one can
compute $p(t)$ and $\vec \phi(t)$ from equations~(\ref{pression})
and~(\ref{field}), respectively; $f(t)$ is then obtained from~(\ref{ef}).

The Einstein-Klein-Gordon equations admit power-law solutions,
$G=t^\alpha$, but they happen to be isotropic and, thus, equal to the ones
discussed in section~\ref{sec:intfrw}. Thus we will analyze, instead, the
general solution of~(\ref{G3}).

Equation~(\ref{G3}) has the first integral 
\begin{equation}
G\frac{\ddot G}{\dot G}+(K-1)\dot G+\frac{M^2}{\dot G}=C,  \label{first}
\end{equation}
where $C$ is an arbitrary constant and 
\begin{equation} 
\label{KM} K\equiv\frac{k^2}{4}-\frac{1}{2},\qquad M^2\equiv
m^2+\frac{a^2}{2}. 
\end{equation}

If instead of $t$ and $G$ we use the new variables $z$ and $\tau$ defined,
for $C\ne0$, in 
\begin{equation} G=z^{1/K}, \qquad t=-\frac{\tau}{C}, 
\label{ansatz} 
\end{equation}
then, equation~(\ref{first}) becomes
\begin{equation} z^{\prime\prime}+z^{-1/K}z^{\prime}+\frac{KM^2}{C^2}
z^{1-2/K}=0, \label{typeq} 
\end{equation}
where a prime denotes the
derivative with respect to $\tau$. This equation is, once more, a
particular case of~(\ref{f}) and can be linearized by using the non-local
transformation~(\ref{3.16}), which in this case is 
\begin{equation}
y\equiv\int{z^{-1/K}\,d z} = K\frac{z^{1-1/K}}{K-1}, \qquad
\eta\equiv\int{ z^{-1/K}\,d \tau} =-\frac{C}{a}p  \label{newxy}
\end{equation}
for $K\ne 1$, and 
\begin{equation}  \label{newp}
y\equiv\int{z^{-1}\,d z} = \ln z, \qquad \eta\equiv\int{z^{-1}\,d \tau}
=- \frac{C}{a}p, 
\end{equation}
for $K=1$. If we take 
\begin{equation}
\beta\equiv(K-1)\frac{M^2}{C^2},\qquad\gamma=0,  \label{beta}
\end{equation}
for $K\ne 1$ and 
\begin{equation}  \label{gamma}
\beta=0,\qquad\gamma =\frac{M^2}{C^2}, 
\end{equation}
for $K=1$,
equation~(\ref{typeq}) reduces to two particular cases of equation
(\ref{3.17}) for $\alpha=1$. The trivial solution of this equation gives
the implicit general solution of (\ref{G3}) which can be written, for
arbitrary $a$, $M$ and non-vanishing $C$, as 
\begin{equation} G=\left[{\rm
e}^{-\eta/2}\left(C_1{\rm e}^{\lambda\eta}+ C_2{\rm e}
^{-\lambda\eta}\right)\right]^{\frac{1}{K-1}}  \label{genimp}
\end{equation}
for $K\ne 1$, and as 
\begin{equation}  \label{gep}
G=C_1e^{-\gamma\eta+C_2e^{-\eta}} 
\end{equation}
for $K=1$. $C_1$ and
$C_2$ are integration constants and $\lambda=\sqrt{ 1-4\beta}/2$.

To check whether a model inflates, we will look at the sign of the
deceleration parameter $q=-\theta^{-2}\left(3\dot\theta+\theta^2\right)$,
where $\theta=u^a_{;a}$ is the expansion and $\dot\theta=\theta_{,a}u^a$,
$ u^a$ being the four-velocity of the cosmic fluid. Since in this case we
are dealing with comoving coordinates, $u^a=\left(e^{-f/2},0,0,0\right)$,
one can see that, apart from a positive factor, the deceleration is
\begin{equation} q\propto 9K\dot G^2-3(2km+C)+(km-C)^2+9M.  \label{eq:q}
\end{equation}

We see from~(\ref{genimp}) that when $K<1$ (i.e., when $k^2<6$) $G$ blows
up for some value
$\eta=\eta_0=\frac{1}{2\lambda}\log\left(-C_2/C_1\right)$, provided that
$C_1C_2<0$. If we expand~(\ref{genimp}) around this value ($
\eta=\eta_0+\delta\eta$) we get 
\begin{equation}
G\propto\delta\eta^{\frac{1}{K-1}}, 
\end{equation}
and from (\ref{ansatz})
and (\ref{newxy}) 
\begin{equation} d\eta=\frac{d\tau}{G}=-\frac{C}{G}dt,
\end{equation}
so that 
\begin{equation} \dot
G=-\frac{C}{G}G^{\prime}\propto \frac{1}{K-1}\delta\eta^{-1}.
\end{equation}
In consequence, when $G\to\infty$ the deceleration
parameter~(\ref{eq:q}) is 
\begin{equation} q\propto
9K\frac{C^2}{(K-1)^2}\frac{1}{\delta\eta^2},\qquad(\mbox{when }
\delta\eta\to0), 
\end{equation}
and there is inflation if $K<0$, i.e., if
$k^2=k_1^2+k_2^2+\cdots+k_n^2<2$. We conclude that in these anisotropic
universes also a greater number of interacting scalar fields makes
inflation less likely.


\subsection{The non-interacting $n$-scalar field problem in the Bianchi
type I model}

\label{sec:nonintbianchi}

We will now assume that the $n$ homogeneous scalar fields $\phi_i$ in the
metric~(\ref{metricz}) do not interact directly, but are driven by a sum
of $ n$ exponential potentials $V_i=V_{0i}{\rm e}^{-k_i\phi_i}$. To
simplify the task of finding exact solutions of the Einstein-Klein-Gordon
equations, we will further assume that $k_1=k_2=\cdots=k_n\equiv k$,
$\phi_1=\phi_2= \cdots=\phi_n\equiv \phi$ and $V_1=V_2=\cdots=V_n\equiv
V=V_0{\rm e} ^{-k\phi} $, so that aforementioned equations can be written
as 
\begin{equation} \dot p=\frac{a}{G},  \label{pression3} 
\end{equation}
\begin{equation} {\rm e}^f=\frac{\ddot G}{2nVG},  \label{ef3}
\end{equation} 
\begin{equation} \frac{\ddot
G}{G}-\frac{1}{2}\left(\frac{\dot G}{G}\right)^2 - \frac{\dot G}{ G}
\dot{f} + \frac{1}{2}\dot p^2 = - n\dot\phi^2,  \label{G23} 
\end{equation}
\begin{equation}  \label{klg3} \ddot{\phi}+\frac{\dot
G}{G}\dot{\phi}-k{\rm e}^f\,V=0, 
\end{equation}
where $a$ is an arbitrary
integration constant. It is easy to check that 
\begin{equation} 
\label{eq:2} \dot{\phi} =\frac{ k}{2n}\frac{\dot G}{G}+\frac{ m}{G}
\end{equation}
is a first integral of the Klein-Gordon
equation~(\ref{klg3}), in terms of the new integration constant $m$.
Inserting (\ref{pression3}) in equation~( \ref{eq:2}) the general
solution of the Klein-Gordon equations is found: 
\begin{equation} 
\label{field3} {\phi}={\phi_0}+p\frac{m}{a}-\frac{k}{2n}\ln G,
\end{equation}
where $\phi_0$ is an arbitrary constant. The equations
(\ref{pression3})--( \ref{G23}) along with (\ref{field3}) uncouple and
their solutions can be obtained if one is able to solve the following
third-order equation for $G$ 
\begin{equation} G\ddot
G^2-{\buildrel\ldots\over G}\dot G G +\left(\frac{1}{2}- \frac{k^2
}{4n}\right) \ddot G\dot G^2 + \left(m^2 +\frac{a^2}{2}\right)\ddot G = 0.
\label{G33} \end{equation}

Since this equation is the same as~(\ref{G3}) once one replaces $
k^2=k_1^2+k_2^2+\cdots+k_n^2$ by $k^2/n$, we may repeat the calculations
of the previous section to reach the opposite conclusion: if several
non-interacting scalar fields are present, they will cooperate to
``assist'' the inflation, which will be more likely and occurs for $k^2
<2n$.


\subsection{Stability of power-law solutions in Bianchi type I model}

\label{sec:stabilityI}

For many purposes it is interesting to investigate the stability of the
solutions of (\ref{G3}). In particular, we hope that the solution
representing an accelerated expansion of the universe, and the solutions
that correspond to the assisted inflation, be stable. To this end we
introduce the variable 
\begin{equation}  \label{o} \Omega=\frac{\dot
h}{h^2}, 
\end{equation}
where $h=\frac{\dot G}{G}$, in equation
(\ref{G3}): 
\begin{equation}  \label{eo}
\dot\Omega+\left[\Omega+K-\frac{M^2}{h^2G^2}\right](\Omega+1)h=0.
\end{equation}

\noindent This equation has the fixed point solution $\Omega=-1$. Note
that equation (\ref{eo}) has also the fixed point solution $\Omega=-K$ if
$\dot G\to\infty$ asymptotically. The corresponding asymptotic limits of
these solutions can be obtained by solving (\ref{o}) for them. The final
result is $G\propto t$ and $G\propto t^{1/K}$ respectively. Let us
investigate the stability of these solutions when $G$ blows up. From
(\ref{eo}) it is easy to see that $\Omega=-1$ is unstable because
expanding the solutions about it, $\Omega=-1+\epsilon$ with $\epsilon \ll
1$, the sign of $\dot\epsilon$ depend on the slope of potential and the
initial conditions. In fact, the corresponding solution $G\propto t$ does
not satisfies Einstein equations (cf.~(\ref{ef3})) and was introduced when
multiplying (\ref{G23}) with $G^2 \ddot{G}$ to obtain (\ref{G33}). On the
other hand, the asymptotic solution $ \Omega=-K$ is stable because the
dynamical equation for the perturbation $ \epsilon$ 
\begin{equation}
\label{ep} \dot\epsilon=-\frac{1-K}{Kt}\,\epsilon 
\end{equation}
near the
attractor indicates that $\epsilon$ decreases for $K<1$. In particular,
the inflationary solutions, that occur for $K<0$, are stable.


\section{The $n$-scalar field problem in a Bianchi~VI$_0$ model}

\label{sec:bianchi6}

The Bianchi~VI$_0$ model can be written as follows: 
\begin{equation}
ds^2 = {\rm e}^{f(t)} \left(-dt^2+dz^2\right) + G(t) \left({\rm
e}^{z}\,dx^2 +{\rm e}^{-z}\,dy^2\right).  \label{metric5} 
\end{equation}

We first consider, as in previous sections, $n$ scalar fields $\phi_i$
interacting through the exponential potential~(\ref{po}). The
corresponding Einstein-Klein-Gordon equations are 
\begin{equation} {\rm
e}^f=\frac{\ddot G}{2VG},  \label{ef5} 
\end{equation} 
\begin{equation}
\frac{\ddot G}{G}-\frac{1}{2}\left(\frac{\dot G}{G}\right)^2 - \frac{\dot
G}{ G} \dot{f} + \frac{1}{2} = - \dot\phi^2,  \label{G25} 
\end{equation}
\begin{equation}  \label{klg5} \ddot{\vec\phi}+\frac{\dot
G}{G}\dot{\vec\phi}-{\rm e}^f\,V\vec k=0. 
\end{equation}
As formerly, one
can check that the vector 
\begin{equation}  \label{eq:25} \dot{\vec\phi}
=\frac{\dot G}{G}\frac{\vec k}{2}+\frac{\vec m}{G}, 
\end{equation}
where
$\vec m$ is a $n$-dimensional arbitrary constant vector, is a first
integral of the Klein-Gordon equation set (\ref{klg5}). By using this
result and the value of $\dot f$ one obtains from (\ref{ef5}), we get
\begin{equation} G\ddot G^2-{\buildrel\ldots\over G}\dot G G
+\left(\frac{1}{2}- \frac{k^2 }{4}\right) \ddot G\dot G^2 + \frac12 \ddot
G G^2+ m^2 \ddot G = 0. \label{G35} 
\end{equation}

\noindent The solutions of this equation are not known. However,
investigating the stability of its fixed points, the asymptotic behavior
of the general solution can be obtained in a simple way. In order to see
whether assisted inflation works in Bianchi type VI$_0$ metrics it is
sufficient to analyze the special case $m^2=0$. In terms of the variable
$ \Omega$ defined by (\ref{o}), equation (\ref{G35}) becomes
\begin{equation}  \label{eo1}
\dot\Omega+\left[\Omega+K-\frac{1}{2h^2}\right](\Omega+1)h=0.
\end{equation}
This equation has three fixed points: $\Omega_1=-1$,
$\Omega_2=-K$ if $ h\to\infty$ asymptotically, and $\Omega_3=0$, which
correspond to $G\propto t $, $G\propto t^{1/K}$ and $G\propto {\rm
e}^{t/\sqrt{2K}}$ respectively. Now, we investigate the stability of these
solutions when $G$ blows up. Expanding the solution about fixed points,
that is, making $ \Omega=\Omega_{1,2,3}+\epsilon$ with $\epsilon \ll 1$,
we get 
\begin{equation}  \label{ep1} \dot\epsilon=\frac{1}{2h}\,\epsilon
\end{equation}
for $\Omega_{1}$, which shows it is unstable, as in the
case discussed in the previous section. On the other hand, one obtains
equation (\ref{ep}) for the linear approximation around $\Omega_2$, and
\begin{equation}  \label{ep3} \dot\epsilon=-\sqrt{\frac{1}{2K}}\epsilon
\end{equation}
in the case of $\Omega_3$. We conclude that the asymptotic
solution $\Omega_2 $ is stable for $K<0$, which means $k^2<2$ (for this
set of potential slopes we have inflation), and $\Omega_3$ is stable for
$K>0$. Note that $G\propto t^{1/K}$ is only an asymptotic solution of
(\ref{G35}), with $m=0$; however, it acts as an attractor for all
solutions that are close to it.

In this special case in which $m^2=0$, one can readily see that the the
deceleration parameter for the solutions $G\propto t^{1/K}$ is (after
recovering the implicit absolute value around $t$) 
\begin{equation}
q\propto \frac{K}{1-K}|t|^{-2-1/K}, 
\end{equation}
which is negative for
$-1/2\le K <0$. Again, the more interacting scalar fields the less likely
is inflation, which ensues when $k^2=k_1^2+k_2^2+ \cdots+k_n^2<2$.

If one assumes now that the scalar fields do not interact directly and one
sets all the fields equal, as in sections~\ref{sec:nonintfrw} and~\ref
{sec:nonintbianchi}, it is easily seen that the attracting power-law
solutions inflate when $k^2<2n$, so that the scalar fields cooperate to
assist inflation.

\section{Conclusions}

We have studied the effects of the appearance of more than one scalar
field both in FRW spacetimes and in anisotropic Bianchi~I cosmologies.
Instead of using the important but particular power-law solutions, we have
taken advantage of the general solution to analyze the generic behavior in
FRW. In Bianchi~I the power-law solutions are isotropic and, thus, very
particular, so that the use of the general solution is even more
illustrative. In all cases we have found, in agreement with calculations
made by other authors with power-law solutions in FRW, that the existence
of more than one scalar field assists inflation provided that they are
uncoupled and interact only through expansion. Also, in this case, the
spectrum of density perturbations becomes closer to scale invariance as
the number of fields increases. If, on the contrary, the fields interact
directly with each other, inflation is less likely to occur. The same
behavior has been obtained in Bianchi~VI$_0$ universes, but (not having
available the general solution) only for power-law solutions, which have
been shown to be attractors when inflation arises. These results reinforce
our belief that the presence of several uncoupled scalar fields in more
general cosmologies fosters inflation, but that, in contradistinction,
mutually interacting scalar fields tend to hinder the inflationary
process.

\section*{Acknowledgments}

This work was partially supported by the University of the Basque Country
through the Research Project~UPV172.310-EB150/98 and the General Research
Grant~UPV172.310-G02/99, as well as by the University of Buenos Aires
under Project TX93. We thank the anonymous referee for his helpful
comments.

\newpage

\section*{Figures}

\bigbreak
\includegraphics[width=\textwidth]{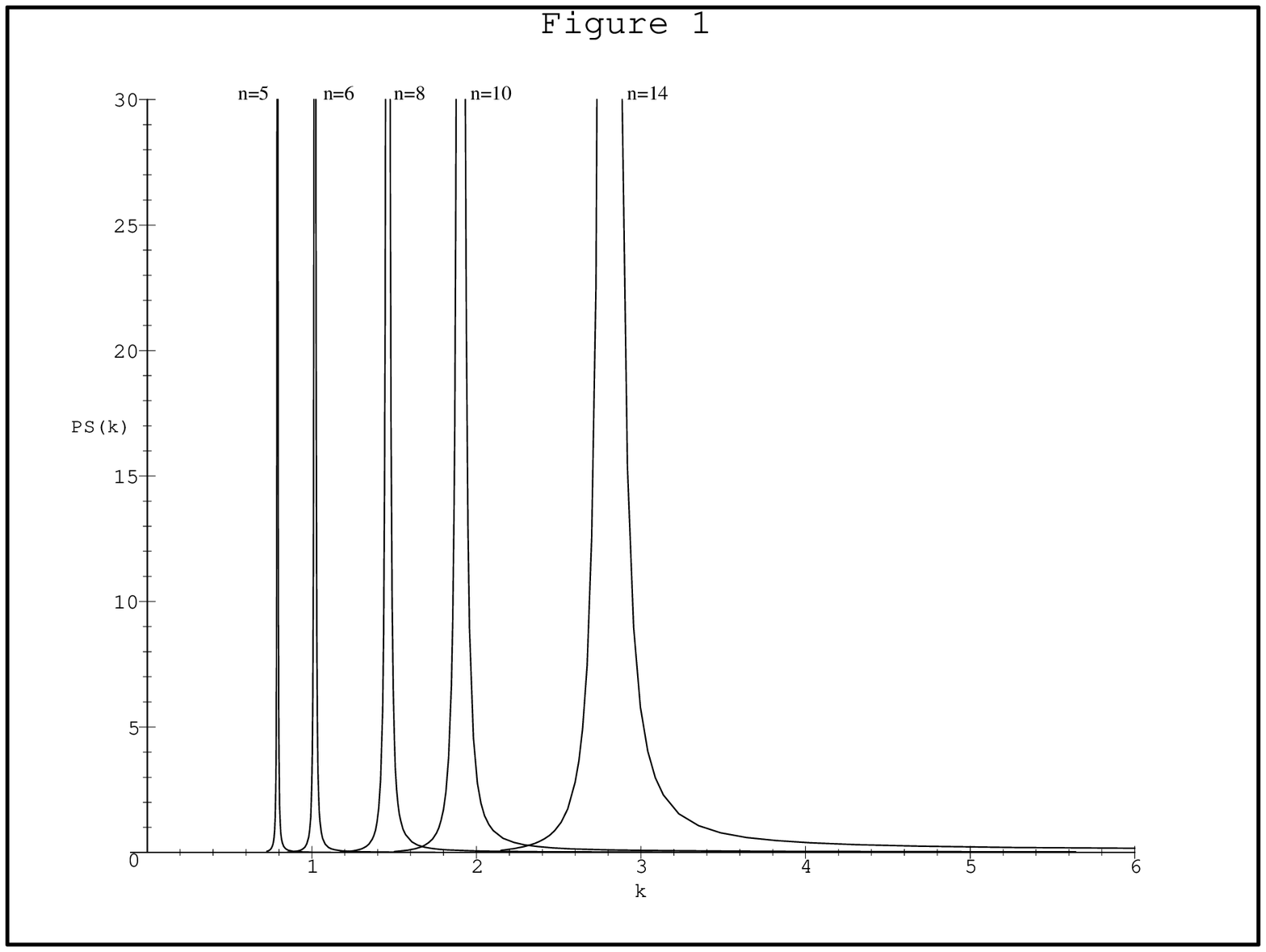}
\medbreak
\noindent
{\bf Figure 1}. Spectrum of the curvature perturbations for $k=3$, $c=-1$,
$ b_1=-1$ and $b_2=0.001$. As $n$ increases (inflation is possible when
$n\ge 5 $) the spectral distribution peak is shifted to high frequencies.

\newpage
\bigbreak
\includegraphics[width=\textwidth]{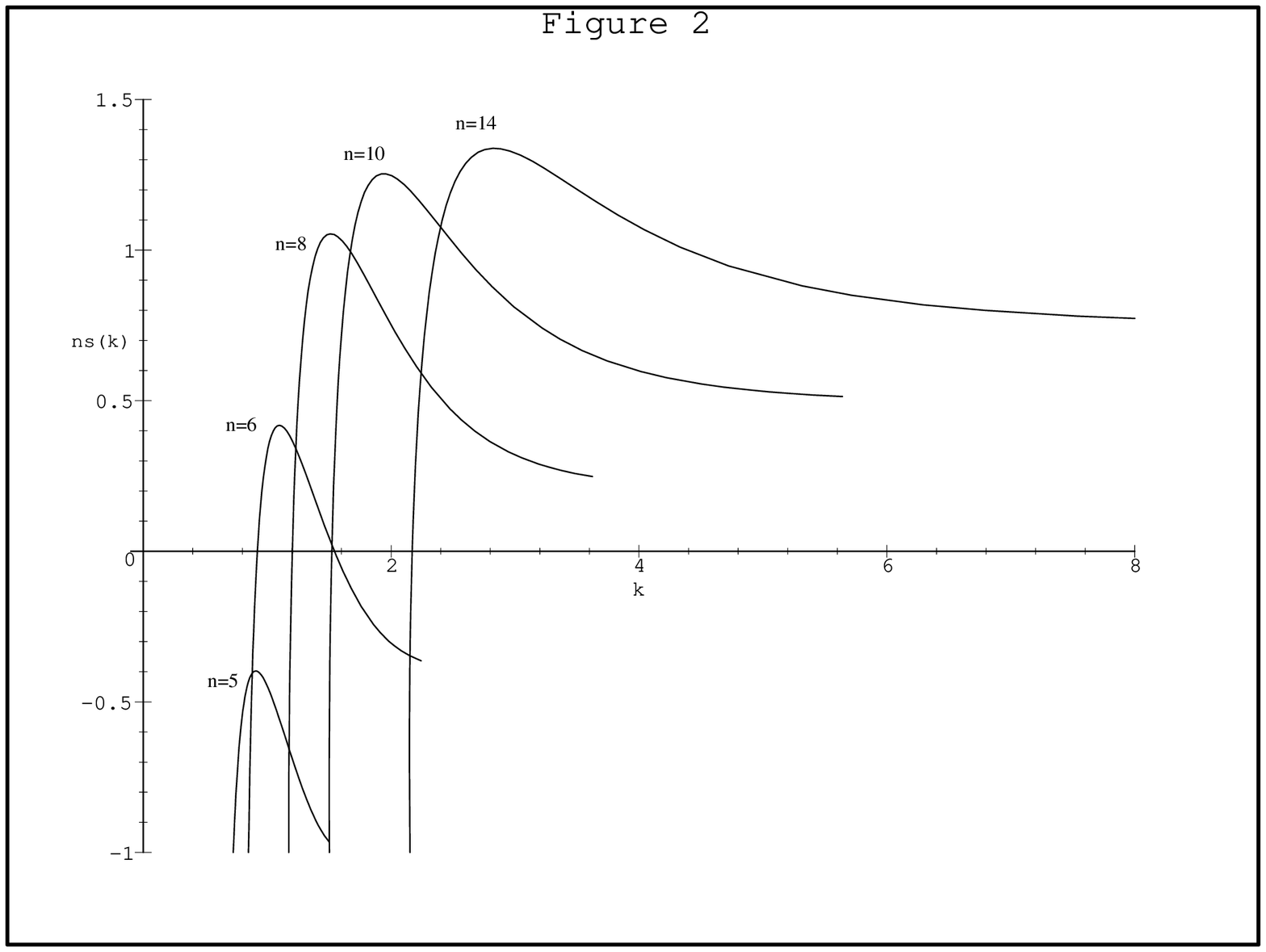}
\medbreak
\noindent
{\bf Figure 2}. The spectral index $n_S({\tilde{k}})$ for the general
solution approaches $1$ in the asymptotic region $a \gg 1$ as the number
of fields increases.

\end{document}